\documentclass[reprint,pra,onecolumn,11pt]{revtex4-2}

\usepackage{graphicx,amsmath,amssymb,}
\usepackage{bm}
\usepackage{scalerel}

\usepackage[top=2cm, bottom=2.5cm, left=2.5cm, right=2.5cm]{geometry}

\begin{document}

\title{Closed form perturbative relativistic modifications to wave-packet dynamics in the quantum harmonic oscillator} 
\author{Jian Carlo Ramos}
\email{jiancarlor@cpp.edu}
\author{Sujoy K. Modak}%
 \email{smodak@cpp.edu}
\affiliation{%
 Department of Physics and Astronomy, California State Polytechnic University, 3801 West Temple Ave., Pomona, CA 91768, USA
}%


\begin{abstract}
We derive closed form expressions of weak relativistic corrections to the wave-packet dynamics of the quantum harmonic oscillator within a perturbative framework. General expressions are derived for the leading-order relativistic contributions to wave-packet parameters, such as the time-dependent widths, variances, and uncertainty relations. Specific calculations are performed for unsqueezed, minimum-uncertainty Gaussian wave packets, valid up to leading order in ($1/c^{2}$). When applied to electron wave packets, the results indicate that relativistic effects become non-negligible for keV-scale harmonic confinement energies: the deviations in variances reach $0.15\% - 1.5\%$ for an electron wave packet confined within the $1-10 \mathrm{keV}$ energy range. We also show that the standard saturation of the uncertainty relation remains unaffected by the leading-order relativistic effect.
\end{abstract}

\maketitle 

\tableofcontents

\section{Introduction}

Since the formulation of quantum mechanics more than a century ago it has gone an enormous number of refinements and additions making it one of the most studied theories in Physics. Precision measurements are becoming increasingly important and central for both the technological developments and foundational inquiry. Modern experiments are pushing for new bounds every day, and it makes many assumptions and approximations of the underlying quantum descriptions to be revisited so that they are aligned with the precision of new technological improvements. This practice would give much more confidence to better analyze the data and draw physical conclusions from them.

Although in most of the laboratory experimental settings we confine ourselves to strong non-relativistic settings, these limits are often pushed upon and we are now within the reach for those experiments where weak relativistic effects are no longer ignorable. It is therefore very important that we accommodate correctly various weak-relativistic modifications  and consider them while drawing physical conclusions from new experimental data. This convergence between the theoretical/mathematical setting and experimental data will be increasingly relevant as new experiments show up in a variety of physical scenarios, from laboratory seetings to observational scenarios. In this work, we investigate one such setting by analyzing weak relativistic corrections to the wave-packet dynamics of the quantum harmonic oscillator and by delineating the parameter regimes in which these corrections become experimentally non-negligible. 

Not surprisingly, there exist a large body of work investigating relativistic extensions of the harmonic oscillator. There are fully relativistic formulations such as the Dirac oscillator introduced by Moshinsky and Szczepaniak~\cite{Moshinsky1989_DiracOsc}, and it subsequent analyses and generalizations~\cite{rel-dir-1,rel-dir-2,rel-dir-3}. There are also works on the Klein--Gordon oscillator \cite{kgo1, kgo2, kgo3, kgo4, kgo6}. These models are exactly solvable and reveal nontrivial relativistic structures such as the spin--orbit couplings, modified spectra, and frequency mixing in wave-packet motion, including Zitterbewegung-type oscillations~\cite{hojo1,hojo2,hojo3,hojo4,hojo5,hojo6}.   Relativistic coherent states and minimum-uncertainty states have also been developed for these models~\cite{relst1, relst2}, providing insight into the interplay of relativity and coherence.  Closer to our work, there are studies on relativistic wave-packet spreading and minimal localization bounds for  gaussian wave packets for relativistic Klein--Gordon or Dirac particles \cite{relwp1,relwp2,relwp3,relwp4,relwp5}.  Although these effects highlight the sensitivity of wave-packet observables to relativistic structure,  they only treat free relativistic particles and do not address bound-state wave-packet evolution in trapping potentials. The latter is the topic for this study but not in a fully relativistic but a weakly relativistic setting.

In the weakly relativistic perturbation theory, one expands the Hamiltonian in powers of $1/c^{2}$ to study relativistic corrections, especially considering the leading Foldy-Wouthuysen (FW) correction, one gets the perturbative part of the Hamiltonian as,
\begin{equation}
H_{\mathrm{rel}}^{(1)} = -\frac{p^{4}}{8 m^{3} c^{2}}.
\label{FW}
\end{equation} 
Recent studies have revisited the quantum harmonic oscillator under weak relativistic corrections to understand how relativistic kinematics subtly reshapes Gaussian wave‑packet evolution. In \cite{relwp5} Huang et al.  analyzed single-particle relativistic Gaussian packets and it was shown that even mild relativistic dispersion induces measurable deviations. Earlier, Guerrero and Aldaya \cite{Guerrero:1998ue} developed a perturbative formalism for the relativistic oscillator, and they demonstrated how the nonlinearity of the relativistic Hamiltonian modifies both frequency spectra and coherent‑state propagation. In another work,  Zarmi \cite{Zarmi2023_RelativisticHO}  further clarified the structure of approximate relativistic eigenstates. 
More recently, Wani et al. \cite{Wani:2025xqd} explored relativistic corrections within the context of quantum‑speed limits for Gaussian systems, underscoring the dynamical constraints imposed by relativistic energy-momentum relations. Collectively speaking, these works have indicated that weak relativistic effects--though perturbative -- can produce distinct signatures in wave‑packet spreading, revival patterns, and semiclassical motion. Therefore, we should consider a deeper examination of controlled relativistic modifications in regimes that are not precisely studied in harmonic oscillator wave-packet dynamics.

In fact, we are not aware of previous work that derive closed-form, time-dependent expressions for the relativistic corrections to the fundamental wave-packet observables for quantum harmonic oscillators. This is indeed a glaring omission and we find it is important that we address this issue here. We calculate new time-dependent expressions for the widths of wave-packets by calculating the time evolution of the Weyl operator in the presence of FW-corrected Hamiltonian.  These modified expressions are then used to calculate expectation values of various time dependent observables which are related to the derivatives of the Weyl operator. Final expressions for modified expectation values of the square of position and momentum operators differ by a compact mathematical form which is expressed as the covariance of specific combinations of relevant operators.  Then we use these modified expressions to specifically study Gaussian shaped wave packets since such wave packets in harmonic potentials provide an essential framework for understanding quantum dynamics, coherent-state evolution, and precision control of trapped particles.  Starting from the FW-corrected Hamiltonian, we derive explicit analytic expressions for the relativistic corrections to the time dependent expressions for position and momentum widths, variances and the uncertainty relationship. These corrections include various harmonic components at $\omega$, $2\omega$, and $3\omega$.

After deriving the explicit modifications for Gaussian wave packets we turn our attention to electron wave packets to numerically estimate these corrections. It may be recalled that for electrons, the non-relativistic harmonic oscillator (HO) describes a vast range of experimental situations with high accuracy; however, as trapping frequencies increase, the relativistic deformation of the kinetic energy induces systematic deviations in wave-packet observables.  The leading FW perturbation \eqref{FW} breaks exact harmonicity and alters the time evolution of the wave-packet width, uncertainty product, and higher moments.  These effects scale with the relativistic Oscillator Quantum to Rest Energy (OQRE) ratio of the wave-packet
\begin{equation}
\eta_E = \frac{\hbar \omega}{m c^{2}}.
\end{equation}
One of the bright points of our study is to show that some of these corrections scale linearly with $\eta_{E}$ and can reach observable levels in next-generation trapped-electron platforms, while remaining within the validity of the $1/c^{2}$ expansion. Indeed, we could show that the modification in variances (both in position and momentum spaces) can reach 0.15\% to 1.5\% level for oscillator energy $\hbar\omega \in [1 keV, 10keV]$, motivating a systematic analysis of relativistic deviations in the dynamics of trapped-electron Gaussian states. On the other hand the leading order relativistic modification to the saturated uncertainty product $\sigma_p\sigma_q = \hbar/2$ only shows up as $1/c^4$ correction and therefore can be safely ignored for the above energy range.  This constitutes, to our knowledge, the first systematic and time-resolved analysis of relativistic modifications to electron wave-packet widths/variances and uncertainty relations in a harmonic potential.

This paper is organized as follows: we provide the mathematical details of calculating observable expectation values considering leading perturbation in section \ref{sec2}. In section \ref{sec3} we use the general results found in \ref{sec2} and calculate them explicitly for Gaussian wave packets. Various modifications for the widths of wave packets and the minimum uncertainty relationships are calculated in this section. In section \ref{sec4} we consider electron wave packets and show that the perturbative relativistic corrections are experimentally relevant for keV scale harmonic confinement energies. Finally, we conclude in section \ref{sec5}. There is one appendix listing various explicit expressions of observable expectation values. 

\section{Weakly Relativistic Quantum Harmonic Oscillator}
\label{sec2}
This section contains the mathematical foundation of our work where we derive main mathematical structures in the perturbative limit of relativistic QHO. These results will then be simplied for the Gaussian wave-packets and numerically calculated for the  electron wave-packets.

\subsection{Perturbative effects on the time evolution of the Weyl Operator}

We start by considering the weakly relativistic harmonic oscillator Hamiltonian
\begin{equation}
\label{hamil}
\hat{H} = \hat{H}_0+\hat{H}^{(1)}_{\mathrm{rel}} = \frac{\hat{p}^2}{2m} + \frac{m\omega^2\hat{q}^2}{2} - \frac{\hat{p}^4}{8m^3c^2}.
\end{equation}
Throughout this paper we work within the perturbative limit to the leading order in $1/c^2$ and neglect $O(c^{-4})$ terms. 

We shall base our work within the statistical operator formalism, for which the state of the system is adequately described by a density operator $\hat\rho$ (which is a positive trace-class operator on the Hilbert space $\mathcal H$ with unit trace), given by
\begin{equation}
\hat\rho=\sum_i P_i |\psi_i\rangle\langle\psi_i|.
\end{equation}
The time  evolution of this operator is governed by the Liouville--von Neumann equation
\begin{equation}
\frac{d}{dt}\hat\rho_t = -\frac{i}{\hbar}[\hat H,\hat\rho_t],
\end{equation}
whose formal solution is the unitary evolution
\begin{equation}
\hat\rho_t=\hat U(t)\hat\rho_0\hat U^\dagger(t),
\qquad
\hat U(t)=e^{-i\hat H t/\hbar}.
\end{equation}
\noindent
Even within this statistical operator formalism, we have a choice to follow either in the Schrödinger picture or  the Heisenberg picture. In fact, it is technically more convenient for us to make use of the dual Heisenberg evolution for bounded operators for which the expectation values satisfy,
\begin{equation}
\mathrm{Tr}(\hat\rho_t\,\hat X)= \mathrm{Tr}(\hat\rho_0\,\hat X_t),
\end{equation}
where the time-dependent operator is
\begin{equation}
\hat X_t = \hat U^\dagger(t)\hat X\hat U(t).
\end{equation}
All observable quantities can be calculated by evolving operators instead of states. 

As the next step, we choose the Weyl operator as generating functional which are defined as \cite{book}
\begin{equation}
\hat W(a,b)= \exp\!\left(\frac{i}{\hbar}(a\hat p+b\hat q)\right).
\end{equation}
The advantage of this is the fact that the position and momentum operators and their higher moments can be directly related with the derivatives of this operator with respect to $a$ and $b$, such that 
\begin{align}
\hat q &= -i\hbar\,\partial_b \hat W(a,b)\big|_{a=b=0},\\
\hat p &= -i\hbar\,\partial_a \hat W(a,b)\big|_{a=b=0}.
\end{align}
Using the Baker-Campbell-Hausdorff identity and $[\hat q,\hat p]=i\hbar$, the Weyl operator can be factorized as follows,
\begin{equation}
\hat W(a,b) = e^{\frac{i}{\hbar}\frac{ab}{2}} e^{\frac{i}{\hbar}a\hat p} e^{\frac{i}{\hbar}b\hat q},
\end{equation}
and  its time evolution is given by
\begin{equation}
\label{Wt}
\hat W_t(a,b) = \hat U^\dagger(t)\hat W(a,b)\hat U(t).
\end{equation}

To treat the relativistic correction perturbatively using the interaction picture, we factorize the unitary evolution operator,
\begin{equation}
\hat U(t) = \hat U_0(t)\hat U_I(t),
\end{equation}
where
\[
\hat U_0(t)=e^{-i\hat H_0 t/\hbar},
\]
and
\begin{equation}
\hat U_I(t) = \tau\left\{\exp\left[ -\frac{i}{\hbar} \int_0^t ds\, \hat U_0^\dagger(s) \hat H^{(1)}_{\mathrm{rel}} \hat U_0(s) \right] \right\}.
\end{equation}
Since $[\hat H_0,\hat H^{(1)}_{\mathrm{rel}}]\neq0$, the perturbation evolves nontrivially in the interaction picture. Truncating the Dyson series \cite{sakurai} to first order in $1/c^2$ yields
\begin{equation}
\label{Ui}
\hat U_I(t) = 1 + \frac{i}{8\hbar m^3c^2}
\int_0^t ds\, e^{\frac{i}{\hbar}\hat H_0 s} \hat p^4 e^{-\frac{i}{\hbar}\hat H_0 s}.
\end{equation}
At this point we define an operator
\begin{equation}\label{4oo}
\mathbb{\hat V}(t) = \int_0^t ds\, e^{\frac{i}{\hbar}\hat H_0 s} \hat p^4 e^{-\frac{i}{\hbar}\hat H_0 s}
\end{equation}
which takes the form 
\begin{equation}\label{4oo2}
\mathbb{\hat V}(t) = \int_0^t ds\, \bigl[\hat p\cos(\omega s)-m\omega\hat q\sin(\omega s)\bigr]^4
\end{equation}
after inserting $\hat U_0^\dagger(s)\hat U_0(s)=1$.

Using the above, to the leading order in $1/c^2$, the time-evolved Weyl operator in \eqref{Wt} becomes
\begin{widetext}
\begin{equation}
\label{Wt_expanded}
\hat W_t^{\text{R}} (a,b) = \hat U_0^\dagger(t)\hat W(a,b)\hat U_0(t) - \frac{i}{8\hbar m^3c^2} \Bigl[ \mathbb{\hat V}(t), \hat U_0^\dagger(t)\hat W(a,b)\hat U_0(t) \Bigr],
\end{equation}
\end{widetext}
where we have introduced a superscript ``R'' to explicitly state that the above expression includes relativistic correction. We shall use this expression for calculating the relativistic corrections to wave-packet dynamics in the next section.

\subsection{Relativistic effects on the position and momentum operators}
By taking the partial derivatives of $W_t(a,b)$ in \eqref{Wt_expanded} with respect to $a$ and $b$, and evaluating at $a=b=0$, we obtain the desired evolutions of the operators. This is shown as follows:\\
\begin{align}
\hat{p}_R (t) &= -i\hbar \frac{\partial}{\partial a}W_t^{\text{R}} (a,b)|_{a=b=0} \label{pt}\\
\hat{q}_R(t)&= -i\hbar \frac{\partial}{\partial b}W_t^{\text{R}} (a,b)|_{a=b=0} \label{qt}\\
\hat{p}_R^2(t) &= -\hbar^2 \frac{\partial^2}{\partial a^2}W_t^{\text{R}} (a,b)|_{a=b=0} \label{p2t}\\
\hat{q}^2_R(t) &= -\hbar^2 \frac{\partial^2}{\partial b^2}W_t^{\text{R}} (a,b)|_{a=b=0} \label{q2t}
\end{align}
For the latter convenience let us first define operator evolution without the relativistic modification. We know that the time evolution of of momentum and position operators for the quantum harmonic oscillator defined by the standard unitary evolution $\hat{p}(t)=U_0^{\dagger}\hat{p}U_0$ and $\hat{q} (t)=U_0^{\dagger}\hat{q}U_0$ satisfy
\begin{equation}\label{nrpq}
\begin{aligned}
\hat{p} (t) &={\hat p} \cos(\omega t)-m\omega{\hat q} \sin(\omega t),\\
\hat{q} (t) &=\frac{{\hat p}}{m\omega}\sin(\omega t)+{\hat q}\cos(\omega t),
\end{aligned}
\end{equation}
where ${\hat p}$ and  ${\hat q}$ are the initial $t=0$ position and momentum operators.

As for the relativistic modification we have to look deeper. First, we need to use our first major result \eqref{Wt_expanded} and then make appropriate derivatives as listed in \eqref{pt}-\eqref{q2t}, to get expressions for time evolution of operators within the relativistic realm. Our next goal is to evaluate derivatives of relativistic commutator brackets appearing in \eqref{Wt_expanded}.  For that we need make use of the following three identities involving the Weyl operator without the relativistic correction: (i) $U_0^{\dagger}W_t(a,b)U_0|_{a=b=0}=1$, (ii) $[\mathbb{\hat V},U_0^{\dagger}W_t(a,b)U_0]_{a=b=0} = 0$, (iii) $\hat{p} (t) = -i\hbar \frac{\partial}{\partial a}W_t(a,b)|_{a=b=0}$  and (iii) $\hat{q}(t) = -i\hbar \frac{\partial}{\partial b}W_t(a,b)|_{a=b=0}$ . Using these simple relations we can also derive various derivatives of various relativistic commutators in \eqref{Wt_expanded}. These are given by
\begin{equation}\label{com1}
    \begin{aligned}
& \frac{\partial}{\partial a}[\mathbb{\hat V},U_0^{\dagger}W(a,b)U_0]_{a=b=0} \\
&= \frac{i}{\hbar}[\mathbb{\hat V},U^{\dagger}_0\hat{p}W(a,b)U_0]_{a=b=0}\\ 
&= \frac{i}{\hbar}\biggr([\mathbb{\hat V},\hat{p}(t)]U_0^{\dagger}W(a,b)U_0+\hat{p}(t)[\mathbb{\hat V},U_0^{\dagger}W(a,b)U_0]\biggr)_{a=b=0} \\
&= \frac{i}{\hbar}[\mathbb{\hat V},\hat{p}(t)],
\end{aligned}
\end{equation}
\begin{equation}\label{com2}
    \begin{aligned}
& \frac{\partial}{\partial b}[\mathbb{\hat V},U_0^{\dagger}W(a,b)U_0]_{a=b=0} \\
&= \frac{i}{\hbar}[\mathbb{\hat V},U_0^{\dagger}\hat{q}W(a,b)U_0]|_{a=b=0}\\
&= \frac{i}{\hbar}\biggr([\mathbb{\hat V},\hat{q}(t)]U_0^{\dagger}W(a,b)U_0+\hat{q}(t)[\mathbb{\hat V},U_0^{\dagger}W(a,b)U_0]\biggr)_{a=b=0} \\
&= \frac{i}{\hbar}[\mathbb{\hat V},\hat{q}(t)],
    \end{aligned}
\end{equation}
\begin{equation}\label{com3}
    \begin{aligned}
& \frac{\partial^2}{\partial a^2}[\mathbb{\hat V},U_0^{\dagger}W(a,b)U_0]_{a=b=0}  \\
&= \frac{i}{\hbar}\biggr(\frac{i}{\hbar}[\mathbb{\hat V},\hat{p}(t)]U_0^{\dagger}\hat{p}W(a,b)U_0+\hat{p}(t)\frac{\partial}{\partial a}[\mathbb{\hat V},U_0^{\dagger}W(a,b)U_0]\biggr)_{a=b=0}\\
&=(\frac{i}{\hbar})^2\biggr(([\mathbb{\hat V},\hat{p}(t)]\hat{p}_s(t)+\hat{p}(t)[\mathbb{\hat V},\hat{p}(t)])U_0^{\dagger}W(a,b)U_0 \\
& +\hat{p}^2(t)[\mathbb{\hat V},U_0^{\dagger}W(a,b)U_0]\biggr)_{a=b=0} \\
&=(\frac{i}{\hbar})^2\biggr([\mathbb{\hat V},\hat{p}^2(t)]U_0^{\dagger}W(a,b)U_0+\hat{p}^2(t)[\mathbb{\hat V},U_0^{\dagger}W(a,b)U_0]\biggr)_{a=b=0} \\
&=(\frac{i}{\hbar})^2[\mathbb{\hat V},\hat{p}^2(t)],
   \end{aligned}
\end{equation}
%
and
\begin{equation}\label{com4}
    \begin{aligned}
& \frac{\partial^2}{\partial b^2}[\mathbb{\hat V},U_0^{\dagger}W(a,b)U_0]_{a=b=0}  \\
&= \frac{i}{\hbar}\biggr(\frac{i}{\hbar}[\mathbb{\hat V},U_0^{\dagger}\hat{q}U_0]U_0^{\dagger}\hat{q}W(a,b)U_0 \\
& +U_0^{\dagger}\hat{q}U_0\frac{\partial}{\partial b}[\mathbb{\hat V},U_0^{\dagger}W(a,b)U_0]\biggr)_{a=b=0} \\
&=(\frac{i}{\hbar})^2\biggr([\mathbb{\hat V},\hat{q}^2(t)]U_0^{\dagger}W(a,b)U_0+\hat{q}^2(t)[\mathbb{\hat V},U_0^{\dagger}W(a,b)U_0]\biggr)_{a=b=0}\\
&=(\frac{i}{\hbar})^2[\mathbb{\hat V},\hat{q}^2(t)].
   \end{aligned}
\end{equation}
%
Now, using the relation \eqref{Wt_expanded} and the above expressions \eqref{com1}-\eqref{com4} in defintions \eqref{pt}-\eqref{q2t}, we can provide the following expressions for the evolution of position/momentum operators in presence of relativistic correction:
\begin{align}
\hat{p}_R(t) &=  \hat{p} (t) - \frac{i}{8\hbar m^3c^2}  [\mathbb{\hat V},\hat{p}(t)], \label{ptr}\\
\hat{q}_R (t) &=   \hat{q} (t) - \frac{i}{8\hbar m^3c^2}  [\mathbb{\hat V},\hat{q}(t)], \label{qtr}\\
\hat{p}^2_R(t) &= \hat{p}^2 (t)  - \frac{i}{8\hbar m^3c^2}  [\mathbb{\hat V},\hat{p}^2(t)], \label{p2tr}\\
\hat{q}^2_R (t) &= \hat{q}^2(t)  - \frac{i}{8\hbar m^3c^2}  [\mathbb{\hat V},\hat{q}^2(t)]. \label{q2tr}
\end{align}
This set of equations is a major preliminary result in this paper which shows how the perturbative fourth order term \eqref{4oo} and \eqref{4oo2} affects the time evolution of the position and momentum operators and their squares.

Our next milestone will be explicitly calculating expressions \eqref{ptr}-\eqref{q2tr}.  For that, first, let us consider the commutators involving the linear terms in momentum and position operators,
\begin{widetext}
\begin{equation}
\begin{aligned}
[\mathbb{\hat V},\hat{p}(t)] &=[\int^t_0 ds(\cos(ws)\hat{p}-mw\sin(ws)\hat{q})^4, \cos(wt)\hat{p}-mw\sin(wt)\hat{q}] \\&= 4i\hbar mw\int^t_0 ds (\cos(ws)\sin(wt)-\sin(ws)\cos(wt))(\cos(ws)\hat{p}-mw\sin(ws)\hat{q})^{3} ,\\
[\mathbb{\hat V},\hat{q}(t)] &=[\int^t_0 ds(\cos(ws)\hat{p}-mw\sin(ws)\hat{q})^4, \frac{\sin(wt)}{mw}\hat{p}+\cos(wt)\hat{q}] \\
& = - 4i\hbar\int^t_0 ds (\sin(ws)\sin(wt)+\cos(ws)\cos(wt))(\cos(ws)\hat{p}-mw\sin(ws)\hat{q})^{3}.
\end{aligned}
\end{equation}
\end{widetext}
To evaluate the right hand side, first we use the expansion 
\begin{widetext}
\begin{equation}
\begin{aligned}
(\cos(ws)\hat{p}-mw\sin(ws)\hat{q})^{3} &= \cos^3(ws)\hat{p}^3 - mw\sin(ws)\cos^2(ws)\big(\hat{p}^{2}\hat{q}
 + \hat{p} \hat{q} \hat{p} + \hat{q} \hat{p}^2\big) \\
 & + m^2w^2\sin^2(ws)\cos(ws) \big(\hat{p}\hat{q}^{2} + \hat{q}\hat{p}\hat{q}+ \hat{q}^2\hat{p}\big)-m^3w^3\sin^3(ws)\hat{q}^3.
 \end{aligned}
\end{equation}
\end{widetext}

At this point we note that for an arbitrary commutator of $p$'s and $q$'s with time-dependent functions $a$, $b$, $\alpha$ and $\beta$, we have $[a\hat{p}+b\hat{q}, \alpha\hat{p}+\beta \hat{q}] = i\hbar(b\alpha-a\beta)$. It also implies,
\begin{equation}
    \begin{aligned}
& [(a\hat{p}+b\hat{q})^n, \alpha\hat{p}+\beta \hat{q}] \\
&= \sum^{n-1}_{j=0} (a\hat{p}+b\hat{q})^j[a\hat{p}+b\hat{q}, \alpha\hat{p}+\beta \hat{q}](a\hat{p}+b\hat{q})^{n-j-1} \\
&= ni\hbar(b\alpha-a\beta) (a\hat{p}+b\hat{q})^{n-1}.
\end{aligned}
\end{equation}

We can now make use of these relevant equations and calculate the relativistic commutators responsible for the perturbing the time evolution of the system. After performing all the integrations and simplifications it leads us to a closed expression for the relativistic commutators involving the linear terms in the position and momentum operators. The first one involves the momentum operator,
\begin{widetext}
\begin{equation}\label{comvp}
\begin{aligned}
[\mathbb{\hat V},\hat{p}(t)] 
&= - i\hbar\{A_{1\omega}(t)\hat{p}^{3} + A_{2\omega}(t)\big(\hat{p}^{2}\hat{q} + \hat{p} \hat{q} \hat{p} + \hat{q} \hat{p}^2\big) +  A_{3\omega}(t)\big(\hat{p}\hat{q}^{2} + \hat{q}\hat{p}\hat{q}+ \hat{q}^2\hat{p}\big) + A_{4\omega}(t)\hat{q}^3\},
\end{aligned}
\end{equation}
\end{widetext}
with the time-dependent functions
\begin{widetext}
\begin{equation}\label{as}
\begin{aligned}
& A_{1\omega}(t) =-\frac{1}{4}m\sin(wt)(6wt+\sin(2wt)),\\
& A_{2\omega}(t) = -\frac{1}{8}m^2w(4wt\cos(wt)-7\sin(wt)+\sin(3wt)),\\
& A_{3\omega}(t) =-\frac{1}{4}m^3w^2\sin(wt)(2wt - \sin(2wt)),\\
&A_{4\omega}(t) = \frac{m^4\omega^3}{8}(-12wt\cos(wt)+9\sin(wt)+\sin(3wt)).
\end{aligned}
\end{equation}
\end{widetext}

Using a similar approach we can also calculate the commutator involving the position operator. This is found to be
\begin{widetext}
\begin{equation}\label{comvq}
\begin{aligned}
[\mathbb{\hat V},\hat{q}(t)] 
& = - i\hbar\{B_{1\omega}(t)\hat{p}^{3} + B_{2\omega}(t)\big(\hat{p}^{2}\hat{q} + \hat{p} \hat{q} \hat{p} + \hat{q} \hat{p}^2\big) + B_{3\omega}(t) \big(\hat{p}\hat{q}^{2} + \hat{q}\hat{p}\hat{q}+ \hat{q}^2\hat{p}\big) + B_{4\omega}(t)\hat{q}^3\},
\end{aligned}
\end{equation}
\end{widetext}
where the new set of time-dependent functions are given by
\begin{widetext}
\begin{equation}\label{bs}
\begin{aligned}
& B_{1\omega}(t) =\frac{1}{8w}(12wt\cos(wt)+11\sin(wt)+3\sin(3wt)), \\
& B_{2\omega}(t) =-\frac{1}{4}m\sin(wt)(2wt+3\sin(2wt)),\\
& B_{3\omega}(t) =\frac{1}{8}m^2w(4wt\cos(wt)+5\sin(wt)-3\sin(3wt)),\\
& B_{4\omega}(t) =-\frac{3m^3\omega^2}{4}\sin(wt)(2wt - \sin(2wt)).
\end{aligned}
\end{equation}
\end{widetext}

Now let us take a moment to analyze more closely our results. It is indeed very interesting to make a crucial observation that tells us something nice about the terms appearing on the right hand sides of \eqref{comvp} and \eqref{comvq} which define some of our major results in \eqref{ptr} and \eqref{qtr}. Let us first introduce the Weyl-ordering rule as a shorthand for the mixed operator terms given by the following formulas
\begin{equation}
\begin{aligned}
\mathcal{W}(q^np^m) &= \frac{1}{2^n}\sum^n_{l=0}\hat{q}^{n-l}\hat{p}^m\hat{q}^l,
\end{aligned}
\end{equation}
\begin{equation}
\begin{aligned}
\mathcal{W}(p^mq^n) &= \frac{1}{2^m}\sum^m_{l=0}\hat{p}^{m-l}\hat{q}^n\hat{p}^l.
\end{aligned}
\end{equation}
Then it is evident that relativistic corrections appearing as the commutator in fact follow a particular pattern in a manner that we can re-express \eqref{comvp} and \eqref{comvq} as
\begin{widetext}
\begin{equation}\label{vps}
\begin{aligned}
& [\mathbb{\hat V}(t),\hat p(t)] = - i\hbar\{A_{1\omega}(t)\hat p^3 +4A_{2\omega}(t)\mathcal W(\hat p^2\hat q) +4A_{3\omega}(t)\mathcal{W}(\hat q^2\hat p)
+A_{4\omega}(t)\hat q^3\}, \\
& [\mathbb{\hat V}(t),\hat q(t)] = - i\hbar\{B_{1\omega}(t)\hat p^3 +4B_{2\omega}(t)\mathcal W(\hat p^2\hat q) +4B_{3\omega}(t)\mathcal{W}(\hat q^2\hat p) +B_{4\omega}(t)\hat q^3\}. \end{aligned}
\end{equation}
\end{widetext}

If we are interested to know the time evolution of the variance/width of the wave-packet as well as the uncertainty relationship in presence of perturbative quartic term due to relativistic dynamics, it is necessary to also calculate time evolution of various second order (quadratic) operators. Now, that we have definite expressions for the relativistic commutators involving linear position and momentum operators, we need to evaluate the commutators involving quadratic terms such as $\hat{p}^2_s(t)$ and $\hat{q}^2_s(t)$. To do that we use the basic algebraic relationship of the commutator algebra
\begin{equation}\label{vs2}
\begin{aligned}
  [\mathbb{\hat V},\hat{p}^2(t)] &= \hat{p}(t)[\mathbb{\hat V},\hat{p}(t)]+[\mathbb{\hat V},\hat{p}(t)]\hat{p}(t),
\end{aligned}
\end{equation}
which relates the quadratic relativistic commutators as the sum of their linear counterparts. This is a crucial identity and we can use it in our next step.

\subsection{Relativistic effects on quantum dynamics}

From the results obtained in last subsection we can readily calculate the time evolution of variances for the wave-packet in the momentum and position spaces with the relativistic correction. For that we make use of explicit expressions derived in \eqref{vps} and \eqref{vs2},  and evaluate the following quantities
\begin{widetext}
\begin{equation}\label{dqs2}
\begin{aligned}
(\sigma_q(t))^2_R &= \langle \hat{q}(t)^2\rangle_{R}-\langle \hat{q}(t) \rangle^2_{R}\\
&=\langle \hat{q}(t)^2 \rangle_{NR} - \frac{i}{8\hbar m^3c^2} \langle [\mathbb{\hat V}(t),\hat{q}(t)^2]\rangle - \langle \hat{q}(t) \rangle^2_{NR} + \frac{i}{4\hbar m^3c^2}\langle \hat{q}(t)\rangle_{NR} \langle [\mathbb{\hat V}(t),\hat{q}(t)]\rangle \\
&=\sigma_q^2 (t)  - \frac{i}{4\hbar m^3c^2} \operatorname{cov} (\hat{q}(t),[\mathbb{\hat V}(t),\hat{q}(t)]),
\end{aligned}
\end{equation}
\end{widetext}
and 
\begin{widetext}
\begin{equation}\label{dps2}
\begin{aligned}
(\sigma_p(t))^2_R&= \langle \hat{p}(t)^2\rangle_{R}-\langle \hat{p}(t) \rangle^2_{R}\\
&=\langle \hat{p}(t)^2 \rangle_{NR} - \frac{i}{8\hbar m^3c^2}  \langle [\mathbb{\hat V}(t),\hat{p}^2(t)]\rangle - \langle \hat{p}(t) \rangle^2_{NR} + \frac{i}{4\hbar m^3c^2}\langle \hat{p}(t)\rangle_{NR} \langle [\mathbb{\hat V}(t),\hat{p}(t)]\rangle\\
&=\sigma_p^2 (t)  - \frac{i}{4\hbar m^3c^2} \operatorname{cov} (\hat{p}(t),[\mathbb{\hat V}(t),\hat{p}(t)]),
\end{aligned}
\end{equation}
\end{widetext}
where we have used the notion of symmetrized covariance between two Hermitian operators $X$ and $Y$ defined by:
\begin{equation}
\begin{aligned}
\operatorname{cov} (X,Y)&= \frac{1}{2}\langle\Delta X \Delta Y + \Delta Y \Delta X \rangle\\
&= \frac{1}{2}(\langle XY\rangle - \langle X\rangle \langle Y\rangle + \langle YX\rangle - \langle Y\rangle \langle X\rangle ) \\
&= \frac{1}{2}(\langle XY\rangle + \langle YX\rangle) -  \langle X\rangle  \langle Y\rangle.
\end{aligned}
\end{equation}
Now we can simplify the covariance terms appearing in \eqref{dqs2} and \eqref{dps2} as
\begin{widetext}
\begin{equation}
\label{covpt}
\begin{aligned}
\operatorname{cov} (\hat{p}(t),[\mathbb{\hat V}(t),\hat{p}(t)])&= \frac{1}{2}\langle [\mathbb{\hat V}(t),\hat{p}(t)^2]\rangle - \langle \hat{p}(t)\rangle\langle [\mathbb{\hat V}(t),\hat{p}(t)]\rangle\\
&= - i\hbar \biggr( A_{1\omega} (t) \operatorname{cov} (\hat{p}(t),\hat{p}^3)+4A_{2\omega}(t) \operatorname{cov} (\hat{p}(t),\mathcal{W}(\hat{p}^2\hat{q}))\\
&+ 4A_{3\omega}(t) \operatorname{cov}  (\hat{p}(t),\mathcal{W}(\hat{q}^2\hat{p}))+A_{4\omega}(t) \operatorname{cov}  (\hat{p}(t),\hat{q}^3)\biggr),\\
\end{aligned}
\end{equation}
\end{widetext}
and
\begin{widetext}
\begin{equation}
\label{covqt}
\begin{aligned}
\operatorname{cov} (\hat{q}(t),[\mathbb{\hat V},\hat{q}(t)])&= \frac{1}{2}\langle [\mathbb{\hat V},\hat{q}(t)^2]\rangle - \langle \hat{q}(t)\rangle\langle [\mathbb{\hat V},\hat{q}(t)]\rangle\\
&= - i\hbar \biggr( B_{1\omega} (t) \operatorname{cov} (\hat{q}(t),\hat{p}^3)+4B_{2\omega}(t) \operatorname{cov} (\hat{q}(t), \mathcal{W}(\hat{p}^2\hat{q}))\\
&+ 4B_{3\omega}(t) \operatorname{cov} (\hat{q}(t), \mathcal{W}(\hat{q}^2\hat{p}))+B_{4\omega}(t) \operatorname{cov}  (\hat{q}(t),\hat{q}^3)\biggr).\\
\end{aligned}
\end{equation}
\end{widetext}
We conclude this section by saying that we now have a complete description of leading relativistic effect on the spreads on wave-packet widths both in the position and momentum spaces. These are given by the equations \eqref{dqs2}, \eqref{dps2}, \eqref{covpt} and \eqref{covqt}. We now move to the next section where we apply these general expressions for coherent gaussian wave-packets.

\section{Relativistic effects on the minimum uncertainty relationship}
\label{sec3}

\subsection{Time preservation of variances and saturation of the uncertainty relation for coherent Gaussian wave-packets}

Since the Heisenberg equations of motion for the harmonic oscillator are linear in $\hat{q}$ and $\hat{p}$, and because the expectation values of those operators satisfy the same classical equations of motion as \eqref{nrpq} (due to Ehrenfest's theorem) the fluctuations evolve by exactly the same linear map as the operators themselves, 
\begin{align}
\Delta \hat p(t)&=\Delta \hat p\cos(\omega t)-m\omega\,\Delta \hat q\sin(\omega t),\\
\Delta \hat q(t)&=\frac{\Delta \hat p}{m\omega}\sin(\omega t)+\Delta \hat q\cos(\omega t).
\end{align}
This is simply the classical symplectic rotation of the phase-space displacement vector $(\Delta\hat{q},\,\Delta\hat{p})$ through angle $\omega t$.

We now consider a coherent, unsqueezed, minimum-uncertainty Gaussian state. The  uncertainty of such a state is distributed isotropically in phase space. Mathematically, such a state can be represented by a wave-packet centered at $q_s$ or $p_s$,
\begin{align*}
\psi (q,t) =  \left(\frac{m\omega}{\pi\hbar}\right)^{1/4}  & \exp\left[-\frac{m\omega}{2\hbar} (q (t) - q_s)^2  + \frac{i p_s (t)}{\hbar}(q (t) - q_s) -\frac{i\omega t}{2} \right]
\end{align*}
and it satisfies the following three concrete conditions
\[
\langle \Delta \hat q\,\Delta \hat p+\Delta \hat p\,\Delta \hat q\rangle=0,
\quad
\sigma_p^2=m^2\omega^2\sigma_q^2,
\quad
\sigma_q\sigma_p=\frac{\hbar}{2}.
\]

Using the relations above, it is easy to show that the position and momentum variances are time-preserved so that  $\sigma_q^2(t)=\sigma_q^2$, $\sigma_p^2(t)=\sigma_p^2$ and 
$\sigma_q(t)\sigma_p(t)=\frac{\hbar}{2}$ for all times. Thus coherent Gaussian wave packets remain minimum-uncertainty states under harmonic evolution. This is a very well-known result within standard Harmonic oscillator Hamiltonian without the quartic relativistic correction considered in this paper.

In the rest of this section we will answer how the inclusion of leading relativistic correction modify these well-known results by invoking a non-symplectic structure to standard classical trajectories.

\subsection{Relativistic corrections to time evolved variances and the uncertainty relationship}
We now ask the same question on the time evolution of variances and uncertainty relationship  in presence of relativistic corrections computed in \eqref{dqs2} and \eqref{dps2} which are modified from their non-relativistic counterparts. Essentially we need to calculate the covariances $cov(\cdot,[\cdot,\cdot])$ for this task. This calculation is rather lengthy and the details are provided in appendix A in section \ref{apa}. The final result, by denoting 
\begin{eqnarray}
a_{1\omega} &=&-m\omega\sin(\omega t), ~~b_{1\omega} = \cos(\omega t) \nonumber\\
a_{2\omega} &=& \cos(\omega t), ~~  b_{2\omega} = \dfrac{\sin(\omega t)}{m \omega}
\end{eqnarray}
is given by,
\begin{equation}
\label{covps}
\operatorname{cov}\!\left(\hat p(t),[\mathbb{\hat V},\hat p(t)]\right) = -i\hbar S_p (t),
\end{equation}
and 
\begin{equation}
\label{covqs}
\operatorname{cov}\!\left(\hat q(t),[\mathbb{\hat V},\hat q(t)]\right) = -i\hbar S_q (t)
\end{equation}
where
\begin{widetext}
\begin{eqnarray}
\label{sp}
S_p (t) &=&  A_{1\omega}(t)b_{1\omega}(t) \left( \frac{3\hbar^4}{16\sigma_q^4} + \frac{3\hbar^2p_s^2}{4\sigma_q^2} \right) +4A_{2\omega}(t) \left\{ b_{1\omega}(t) \left( \frac{3\hbar^2p_s q_s}{8\sigma_q^2} \right) + a_{1\omega}(t) \left( \frac{3\hbar^2}{16} + \frac{3}{4}p_s^2\sigma_q^2 \right) \right\}\nonumber\\
&& +4A_{3\omega}(t)
\left\{
b_{1\omega}(t)
\left(
\frac{3\hbar^2}{16\sigma_q^2}
(q_s^2+\sigma_q^2)
\right)
+
a_{1\omega}(t)
\left(
\frac{3p_sq_s\sigma_q^2}{2}
\right)
\right\} +A_{4\omega}(t)a_{1\omega}(t)
\{
3\sigma_q^2(q_s^2+\sigma_q^2)
\} . \nonumber\\
\end{eqnarray}
and
\begin{eqnarray}
\label{sq}
S_q (t) &=&  B_{1\omega}(t)b_{2\omega}(t)\left( \frac{3\hbar^4}{16\sigma_q^4} + \frac{3\hbar^2p_s^2}{4\sigma_q^2} \right) +4B_{2\omega}(t)
\left\{ b_{2\omega}(t) \left( \frac{3\hbar^2p_sq_s}{8\sigma_q^2} \right) + a_{2\omega}(t) \left( \frac{3\hbar^2}{16} + \frac{3}{4}p_s^2\sigma_q^2 \right) \right\}
\nonumber\\
&& +4B_{3\omega}(t)
\left\{
b_{2\omega}(t)
\left(
\frac{3\hbar^2}{16\sigma_q^2}
(q_s^2+\sigma_q^2)
\right)
+
a_{2\omega}(t)
\left(
\frac{3p_sq_s\sigma_q^2}{2}
\right)
\right\} 
 +B_{4\omega}(t)a_{2\omega}(t)
\{
3\sigma_q^2(q_s^2+\sigma_q^2)
\}. \nonumber\\
\end{eqnarray}
\end{widetext}
\vspace{0.5cm}
Substituting \eqref{covps} and \eqref{covqs} in \eqref{dps2} and \eqref{dqs2} respectively, we arrive at the following results 
\begin{equation}
\label{relpv}
(\sigma_p(t))^2_R = \sigma_p^2 - \frac{1}{4m^3 c^2} S_p(t)
\end{equation}
and 
\begin{equation}
\label{relqv}
(\sigma_{q}(t))^2_R = \sigma_q^{2} - \frac{1}{4m^3 c^2} S_q(t)
\end{equation}
These expressions show the time evolution of the position and momentum space variances with the leading relativistic corrections. Notice that these are not time-preserved -- deviating from the strict non-relativistic case. The novelty of this work is calculating these semi-relativistic modifications explicitly and in a closed form.

We are now in a position to make use of the above equations and formally write down the modified uncertainty relationship in presence of quartic (in momentum) term in otherwise HO Hamiltonian. For that, let us consider the equations \eqref{dqs2} and \eqref{dps2} and use them to to evaluate the following algebraic relationship
\begin{widetext}
\begin{equation}
\begin{aligned}
(\sigma_q^2 (t) \sigma_p^2 (t))_R = \sigma_q^2 (t) \sigma_p^2 (t) &- \frac{i}{4\hbar m^3c^2} \{ \sigma_p^2 \operatorname{cov} (\hat{q}(t),[\hat{\mathbb{V}},\hat{q}(t)]) +\sigma_q^2  \operatorname{cov} (\hat{p}(t),[\hat{\mathbb{V}},\hat{p}(t)])\} 
\end{aligned}
\end{equation}
\end{widetext}
To make the algebra trackable and  to filter the most dominant relativistic effect we now further simplify the above expression to obtain, 
\begin{equation}
\left(\sigma_q(t)\sigma_p(t)\right)_R
=
\sigma_q(t)\sigma_p(t)
\sqrt{
1
-
\frac{1}{4m^3c^2}
\left[
\frac{S_q(t)}{\sigma_q^2(t)}
+
\frac{S_p(t)}{\sigma_p^2(t)}
\right]
}
\end{equation}
Thus even for the unsqueezed gaussian wave-packet $(\sigma_q (t) \sigma_p (t))_R \neq \frac{\hbar}{2}$ in presence of the relativistic effects. The time-preserved structure is no-longer valid and we must use the following relationship which is true up to the leading order in relativistic correction
\begin{equation}\label{relmug}
\left(\sigma_q\sigma_p\right)_R
\approx
\frac{\hbar}{2}
-
\frac{\hbar}{16m^3c^2}
\left[
\frac{S_q(t)}{\sigma_q^2}
+
\frac{4\sigma_q^2}{\hbar^2}S_p(t)
\right].
\end{equation}
This is another main result of this paper which show how perturbative relativistic effects modify the time-preserved dynamics of unsqueezed gaussian wave-packet  and thereby provide us a new lower bound of uncertainty relationship. We can now ask a more practical question which is related to its measurable implications in a laboratory setting. It turns out that electron wave-packets stands out as the most promising candidate for most relevant numerical estimations. We shall delve into this issue in the next section.

\section{Numerical Estimation of Relativistic Corrections to Wave-Packet Dynamics}
\label{sec4}

We analyze here the leading $1/c^{2}$ relativistic corrections to the dynamics of an Gaussian wave packet of mass $m$ evolving in a harmonic potential of frequency $\omega$.  Both the position--momentum uncertainty product and the packet width acquire corrections governed by the dimensionless parameter
\begin{equation}
\eta_{E} \equiv \frac{\hbar\omega}{m c^{2}},
\label{etae}
\end{equation}
which is the {Oscillator Quantum to the Rest Energy} (OQRE) ratio of the wave-packet under consideration. Throughout, we consider times of order a few oscillation periods ($t\sim \mathcal{O}(1/\omega)$), for which the first-order relativistic expansion is well controlled.

First, let us estimate the modifications to variance and widths as defined in equations \eqref{relpv} and \eqref{relqv}. We shall consider a near-ground-state centered Gaussian packet, for which we can set
\begin{equation}
p_{s}\approx 0, \qquad q_{s}\approx 0, \qquad
\sigma_{s}^{2}\simeq \frac{\hbar}{2m\omega},
\label{Grsp}
\end{equation}
and all occurrences of $\sigma_{0}$ may be eliminated in favor of $\hbar/(m\omega)$. By substituting this in \eqref{sp} and \eqref{sq} and simplifying we obtain,
\begin{eqnarray}
S_p(t) &=& - \frac{3\hbar^2}{2}  m^3 \omega^2 \sin ^2(\omega t) \\
S_q (t) &=&  \frac{3 \hbar^2}{2}  m \sin^2(\omega t).
\end{eqnarray}
Substituting these expressions in \eqref{relpv} and \eqref{relqv}, we can readily calculate
\begin{eqnarray}
\frac{(\sigma_p^2)_R - \sigma_p^2}{\sigma_p^2} = \frac{3}{4} \eta_{E} \sin^2(\omega t) \label{keyr}\\
\frac{(\sigma_q^2)_R - \sigma_q^2}{\sigma_q^2} = - \frac{3}{4} \eta_{E} \sin^2(\omega t) \label{keyr2}.
\end{eqnarray}
Therefore the maximum variance modification occur when $\sin(\omega t)=1$, and this is given by the following elegant equation,
\begin{eqnarray}
\left| \frac{\Delta \sigma_p^2}{ \sigma_p^2} \right|_{\text{max}} = \left| \frac{\Delta \sigma_q^2}{ \sigma_q^2} \right| = \frac{3}{4} \eta_{E}.
\end{eqnarray}
On the other hand, if we only want to measure the deviation in width of the wave-packet, we find
\begin{eqnarray}
\left| \frac{\Delta \sigma_p}{ \sigma_p} \right|_{\text{max}} = \left| \frac{\Delta \sigma_q}{ \sigma_q} \right| \simeq \frac{3}{8} \eta_{E}.
\end{eqnarray}
Now, we can also ask for the change in uncertainty relationship modification. For that we just need to use \eqref{keyr} and \eqref{keyr2}, and derive the following
\begin{eqnarray}
(\sigma_p (t) \sigma_q (t) )_R &=& \sigma_p \sigma_q \sqrt{1-\frac{9}{16}\eta_E^2\sin^2(\omega t)} \\
                                             &=& \sigma_p \sigma_q + {\cal{O}} (1/c^4).
\end{eqnarray}
We can therefore conclude that although the width/variance of the wave-packet is affected by the leading order $({\cal{O}} (1/c^2))$ relativistic correction, the uncertainty relationship remains unchanged to this order. The leading modifications to the uncertainty for an unsqueezed wave-packet comes only at the ${\cal{O}} (1/c^4)$ order which can be safely neglected.
Also, our results are valid for any wave-packets beyond the electron wavepecket we will consider here -- all we need to use the appropriate mass in the definition of OQRE ratio in \eqref{etae}.

We can now make a numerical estimate of these modifications considering the rest mass energy of electron $m_e c^2 = 511$ keV. For $\hbar\omega = 1$ keV, we get $\eta_E = 1.9569 \times 10^{-3}$, and for $\hbar\omega = 10$ keV its value is $\eta_E = 1.9569 \times 10^{-2}$. For this range of the quantum oscillator energy, we can estimate $$\left| \frac{\Delta \sigma_p^2}{ \sigma_p^2} \right|_{\text{max}} =\left| \frac{\Delta \sigma_q^2}{ \sigma_q^2} \right|_{\text{max}} \in [0.1468, 1.468],$$ i.e., the modifications lie roughly between the range $0.15\%$ to $1.5\%$. This opens up a possibility of examining these deviations in a realistic experimental set up with present/near future technologies.. However, the modifications to the uncertainty relationship will be extremely small and can be easily ignored while treating perturbative relativistic corrections to wave-packet dynamics.

\section{Summary and Conclusions}
\label{sec5}
In this work we derived  closed-form, time-dependent expressions considering leading relativistic corrections to the fundamental wave-packet observables for quantum harmonic oscillators for the very first time. We used the density matrix formalism and studied time evolution of the Weyl operator with leading relativistic correction. We tracked the modifications in expectation values of position and momentum operators as well as their squared versions.  These perturbative corrections to the dynamical equations were manifested as the covariance of specific combinations of position and momentum operators and their products. This compact mathematical form was explicitly calculated for Gaussian shaped wave-packets which are essential for our understanding of quantum dynamics, coherent state behavior and precision control of trapped particles. 

We provided closed form expressions for the time-dependent width in position and momentum spaces considering the leading relativistic correction ($1/c^2$) in general. Appearance of higher harmonics (up to 3rd overtone) were all due to relativistic effects. These expressions were explicitly simplified for unsqueezed Gaussian wave-packets which show the following interesting result: although the time-dependent widths, variances were perturbed in $1/c^2$  order of magnitude, the  closed expression for the relativistically modified uncertainty relationship inherited only $1/c^4$ modification which can be safely ignored. Finally, we made a numerical estimate of the leading order weak relativistic contributions  for electron wave-packets where electrons stay in harmonic traps of strengths between 1 to 10 keV.  These deviations stayed in the range between 0.15\% to 1.5\%  which might be experimentally measured  if we can trap electrons in keV scale potentials. This work provides us a platform for a future study where we plan to further scrutiny the ramifications of our results in other types of precision experiments  across various active research fields.

\begin{acknowledgments}
SKM thanks the California State Polytechnic University Pomona, for financial support through the faculty start up fund.
\end{acknowledgments}


\section{Appendix A}
\label{apa}
In this appendix we provide simplified expressions of various mathematical identities and their explicit calculations considering Gaussian wave-packets. 
First, the definition of covariance between the position and momentum operators used in this work (Section II) are the following:
\begin{equation}
\begin{aligned}
\operatorname{cov} (\hat{p},\hat{p}^3) &= \langle\hat{p}^4\rangle-\langle\hat{p}\rangle\langle\hat{p}^3\rangle,\\
\operatorname{cov}(\hat{q},\hat{q}^3) &= \langle\hat{q}^4\rangle-\langle\hat{q}\rangle\langle\hat{q}^3\rangle,\\
\operatorname{cov}(\hat{p},\hat{q}^3) &= \frac{1}{2}(\langle \hat{p}\hat{q}^3\rangle+\langle \hat{q}^3\hat{p}\rangle)-\langle\hat{p}\rangle\langle\hat{q}^3\rangle, \\
\operatorname{cov}(\hat{q},\hat{p}^3) &= \frac{1}{2}(\langle \hat{q}\hat{p}^3\rangle+\langle \hat{p}^3\hat{q}\rangle)-\langle\hat{q}\rangle\langle\hat{p}^3\rangle. \\
\end{aligned}
\end{equation}
Next, the definition various covariances between the position/momentum operator with Weyl ordered operator combinations,
\begin{equation}
\begin{aligned}
\operatorname{cov}(\hat{p},\mathcal{W}(\hat{p}^2\hat{q})) &= \frac{1}{2}(\langle \hat{p}\mathcal{W}(\hat{p}^2\hat{q})\rangle+\langle \mathcal{W}(\hat{p}^2\hat{q})\hat{p}\rangle)-\langle\hat{p}\rangle\langle \mathcal{W}(\hat{p}^2\hat{q})\rangle ,\\
\operatorname{cov}(\hat{q},\mathcal{W}(\hat{p}^2\hat{q})) &= \frac{1}{2}(\langle \hat{q}\mathcal{W}(\hat{p}^2\hat{q})\rangle+\langle \mathcal{W}(\hat{p}^2\hat{q})\hat{q}\rangle)-\langle\hat{q}\rangle\langle \mathcal{W}(\hat{p}^2\hat{q})\rangle ,\\
\operatorname{cov}(\hat{p},\mathcal{W}(\hat{q}^2\hat{p})) &= \frac{1}{2}(\langle \hat{p}\mathcal{W}(\hat{q}^2\hat{p})\rangle+\langle \mathcal{W}(\hat{q}^2\hat{p})\hat{p}\rangle)-\langle\hat{p}\rangle\langle \mathcal{W}(\hat{q}^2\hat{p})\rangle ,\\
\operatorname{cov}(\hat{q},\mathcal{W}(\hat{q}^2\hat{p})) &= \frac{1}{2}(\langle \hat{q}\mathcal{W}(\hat{q}^2\hat{p})\rangle+\langle \mathcal{W}(\hat{q}^2\hat{p})\hat{q}\rangle)-\langle\hat{q}\rangle\langle \mathcal{W}(\hat{q}^2\hat{p})\rangle .
\end{aligned}
\end{equation}
%
All of the above operator orderings can be explicitly calculated using the Gaussian wave-packets defined in Section III.B. The relevant expectation values are readily obtained as:
\begin{equation}
\begin{aligned}
\langle\hat{q}\rangle &= q_s,\\
\langle\hat{q}^3\rangle &=  q_s^3+3q_s\sigma_q^2,\\
\langle\hat{q}^4\rangle &=  q_s^4+6q_s^2\sigma_q^2+3\sigma_q^4,\\
\langle\hat{p}\rangle &= p_s ,\\
\langle\hat{p}^3\rangle &=  \frac{3\hbar^2p_s}{4\sigma_q^2}+p_s^3, \\
\langle\hat{p}^4\rangle &= \frac{3\hbar^4}{16\sigma_q^4} +\frac{3}{2\sigma_q^2}\hbar^2p_s^2+p_s^4. \\
\langle\hat{p}\hat{q}^3\rangle &= p_s(q_s^3+3q_s\sigma_q^2)-\frac{3}{2}i\hbar(q_s^2+\sigma_q^2),\\
\langle\hat{q}^3\hat{p}\rangle &= p_s(q_s^3+3q_s\sigma_q^2)+\frac{3}{2}i\hbar(q_s^2+\sigma_q^2),\\
\langle\hat{q}\hat{p}^3\rangle &= \frac{8p_s^3q_s\sigma_q^2+ 12i\hbar p_s^2\sigma_q^2+6\hbar^2 p_sq_s+3i\hbar^3}{8\sigma_q^2},\\
\langle\hat{p}^3\hat{q}\rangle &= \frac{8p_s^3q_s\sigma_q^2-12i\hbar p_s^2\sigma_q^2+6\hbar^2 p_sq_s-3i\hbar^3}{8\sigma_q^2}.\\
\end{aligned}
\end{equation}
In addition, we can show that
\begin{equation}
\begin{aligned}
\langle \mathcal{W}(\hat{p}^2\hat{q})\rangle &= \frac{3q_s(4 p_s^2\sigma_q^2+\hbar^2)}{16\sigma_q^2},\\
\langle\hat{q}\mathcal{W}(\hat{p}^2\hat{q})\rangle &= \frac{3(4p_s^2\sigma_q^2(q_s+\sigma_q^2)+4i\hbar p_sq_s\sigma_q^2+\hbar^2(q_s^2+\sigma_q^2))}{16\sigma_q^2}, \\
\langle \mathcal{W}(\hat{p}^2\hat{q})\hat{q}\rangle &= \frac{3(4p_s^2\sigma_q^2(q_s+\sigma_q^2)-4i\hbar p_sq_s\sigma_q^2+\hbar^2(q_s^2+\sigma_q^2))}{16\sigma_q^2} ,\\
\langle \mathcal{W}(\hat{p}^2\hat{q})\hat{p}\rangle &= \frac{3(8p_s^3q_s\sigma_q^2+4i\hbar p_s\sigma_q^2+6p_sq_s\hbar^2+i\hbar^3)}{32\sigma_q^2}, \\
\langle\hat{p}\mathcal{W}(\hat{p}^2\hat{q})\rangle &= \frac{3(8p_s^3q_s\sigma_q^2-4i\hbar p_s\sigma_q^2+6p_sq_s\hbar^2-i\hbar^3)}{32\sigma_q^2},
\end{aligned}
\end{equation}
\begin{equation}
\begin{aligned}
\langle \mathcal{W}(\hat{q}^2\hat{p})\rangle &= \frac{3p_s(q_s^2+\sigma_q^2)}{4},\\
\langle\hat{q}\mathcal{W}(\hat{q}^2\hat{p})\rangle &= \frac{3}{8}(2p_s(q_s^3+3q_s\sigma_q^2)+i\hbar(q_s^2+\sigma_q^2)),\\
\langle \mathcal{W}(\hat{q}^2\hat{p})\hat{q}\rangle &= \frac{3}{8}(2p_s(q_s^3+3q_s\sigma_q^2)-i\hbar(q_s^2+\sigma_q^2)),\\
\langle\hat{p}\mathcal{W}(\hat{q}^2\hat{p})\rangle &= \frac{3(4p_s^2\sigma_q^2(q_s+\sigma_q^2)-4i\hbar p_sq_s\sigma_q^2+\hbar^2(q_s^2+\sigma_q^2))}{16\sigma_q^2}, \\
\langle \mathcal{W}(\hat{q}^2\hat{p})\hat{p}\rangle &= \frac{3(4p_s^2\sigma_q^2(q_s+\sigma_q^2)+4i\hbar p_sq_s\sigma_q^2+\hbar^2(q_s^2+\sigma_q^2))}{16\sigma_q^2}. 
\end{aligned}
\end{equation}
Using the results above, relevant covariance terms for the time-independent operators are obtained as follows:
\begin{equation}
\begin{aligned}
\operatorname{cov}(\hat{q},\hat{q}^3) &= 3\sigma_q^2(q_s^2+\sigma_q^2), \\
\operatorname{cov}(\hat{p},\hat{p}^3) &=\frac{3\hbar^4}{16\sigma_q^4}+\frac{3\hbar^2}{4\sigma_q^2}p_s^2,\\
\operatorname{cov}(\hat{p},\hat{q}^3) &= 0,\\
\operatorname{cov}(\hat{q},\hat{p}^3) &= 0,
\end{aligned}
\end{equation}
and
\begin{equation}
\begin{aligned}
\operatorname{cov}(\hat{p},\mathcal{W}(\hat{p}^2\hat{q})) &= \frac{3\hbar^2p_sq_s}{8\sigma_q^2},\\
\operatorname{cov}(\hat{q},\mathcal{W}(\hat{p}^2\hat{q})) &=\frac{3\hbar^2}{16} + \frac{3}{4} p_s^2 \sigma_q^2,\\
\operatorname{cov}(\hat{p},\mathcal{W}(\hat{q}^2\hat{p})) &= \frac{3\hbar^2}{16\sigma_q^2}(q_s^2+\sigma_q^2),\\
\operatorname{cov}(\hat{q},\mathcal{W}(\hat{q}^2\hat{p})) &= \frac{3p_sq_s\sigma_q^2}{2}.
\end{aligned}
\end{equation}
These terms are then used to calculate the key expressions in \eqref{sp} and \eqref{sq}.




\begin{thebibliography}{99}

\bibitem{Moshinsky1989_DiracOsc}
M.~Moshinsky and A.~Szczepaniak, ``The Dirac oscillator,'' J.\ Phys.\ A \textbf{22}, L817 (1989).

\bibitem{rel-dir-1}
J. Bentez, R. P. Martnez y Romero, H. N. Núez-Yépez, and A. L. Salas-Brito, ``Solution and hidden supersymmetry of a Dirac oscillator'', Phys. Rev. Lett. 64, 1643 (1990); Erratum Phys. Rev. Lett. 65, 2085 (1990)

\bibitem{rel-dir-2}
R.~P.~Martinez-y-Romero, H.~N.~Nunez-Yepez and A.~L.~Salas-Brito, ``Relativistic quantum mechanics of a Dirac oscillator,''
Eur. J. Phys. \textbf{16}, 135-141 (1995).

\bibitem{rel-dir-3}
M. Moshinsky, \& E. Sadurn\'i,  ``Time dependent problems in relativistic quantum mechanics'', Revista Mexicana de Fisica Supplement. 52 (2006). 

\bibitem{kgo1} S. Bruce, P. Minning, ``The Klein-Gordon oscillator,'' Nuov Cim A 106, 711–713 (1993). 

\bibitem{kgo2}
V.~V.~Dvoeglazov, ``Comment on `The Klein-Gordon oscillator' (S. Bruce and P. Minning, Nuovo Cim. 106A (1993) 711),'' Nuovo Cim. A \textbf{107}, 1413-1418 (1994)

\bibitem{kgo3}
Y. Nedjadi and R. C. Barrett, ``The Duffin-Kemmer-Petiau oscillator'',  J. Phys. A: Math. Gen. 27 4301 (1994).
\bibitem{kgo4}
Y Nedjadi and R C Barrett, ``A generalized Duffin-Kemmer-Petiau oscillator,''  J. Phys. A: Math. Gen. 31 6717 (1998).

\bibitem{kgo6}
M.~G.~Garcia, A.~S.~de Castro, L.~B.~Castro and P.~Alberto,
 ``New solutions of the D -dimensional Klein{\textendash}Gordon equation via mapping onto the nonrelativistic one-dimensional Morse potential,'' Annals Phys. \textbf{378}, 88-99 (2017)


\bibitem{hojo1} A.O. Barut and A. J. and Bracken, ``Zitterbewegung and the internal geometry of the electron,'' Phys. Rev. D 23, 2454 (1981).

\bibitem{hojo2}
K.Y. Bliokh et al. ``Relativistic spin-orbit interactions of photons and electrons,''Physical Review A, 97, 043840 (2018).

\bibitem{hojo3}
S. Lovett, P.M. Walker, A. Osipov et al., ``Observation of Zitterbewegung in photonic microcavities,'' Light Sci Appl 12, 126 (2023). 

\bibitem{hojo4}
K. Y. Bliokh, M. R. Dennis, and F. Nori, ``Relativistic Electron Vortex Beams: Angular Momentum and Spin-Orbit Interaction,'' Phys. Rev. Lett. 107, 174802 (2011).

\bibitem{hojo5}
K. Bliokh, F. Rodr\'iguez-Fortu\~no, F. Nori et al. ``Spin–orbit interactions of light,'' Nature Photon 9, 796–808 (2015). 

\bibitem{hojo6}
A. Mukherjee and  S. Sengupta, ``Phase mixing of relativistically intense longitudinal wave packets in a cold plasma,'' Phys. Plasmas 23, 092112 (2016).








\bibitem{relst1}K. Kowalski, J. Rembieli\'nski, J. P. Gazeau, ``On the coherent states for a relativistic scalar particle,'' Annals of Physics, 399, 204-223 (2018).

\bibitem{relst2}
A.~Mostafazadeh and F.~Zamani, ``Quantum mechanics of Klein-Gordon fields. II. Relativistic coherent states,''
Annals Phys. \textbf{321}, 2210-2241 (2006)



\bibitem{relwp1}T. D. Newton and E. P. Wigner, ``Localized States for Elementary Systems,'' Rev. Mod. Phys. 21, 400 (1949). 

\bibitem{relwp2}C. Almeida  A. Jabs, ``Spreading of a relativistic wave packet,'' Am. J. Phys. 52, 921–925 (1984).

\bibitem{relwp3} Hung-Ming Tsai and Bill Poirier, ``Exploring the propagation of relativistic quantum wave-packets in the trajectory-based formulation,''  J. Phys.: Conf. Ser. 701 012013 (2016).

\bibitem{relwp4} C. Anastopoulos and N. Savvidou, ``Time of arrival and localization of relativistic particles,''  J. Math. Phys. 60, 032301 (2019).

\bibitem{relwp5}
Y.~C.~Huang, F.~M.~He and S.~Y.~Lin, ``Quantum mechanical Gaussian wave-packets of single relativistic particles,'' Chin. J. Phys. \textbf{87}, 486-495 (2024).




\bibitem{Guerrero:1998ue}
J.~Guerrero and V.~Aldaya, ``A Perturbative approach to the relativistic harmonic oscillator,'' Mod. Phys. Lett. A \textbf{14}, 1689-1699 (1999)


\bibitem{Zarmi2023_RelativisticHO}
Y. Zarmi, ``On the Relativistic Harmonic Oscillator,'' Applied Mathematics , 14, 1-20 (2023).



\bibitem{Wani:2025xqd}
S.~S.~Wani, A.~K.~Khan, S.~Al-Kuwari and M.~Faizal,
 ``Relativistic quantum-speed limit for Gaussian systems and prospective experimental verification,'' Phys. Lett. A \textbf{565}, 131147 (2026).




\bibitem{book} Alexander Holevo, ``Probabilistic and Statistical Aspects of Quantum Theory,'' Quaderni della Normale; Edizioni della Normale: Pisa, Italy, 2011.
\bibitem{sakurai} J.J. Sakurai, J Napolitano, ``Modern Quantum Mechanics,'' Chapter 5, Third Edition, 2021. 




\end{thebibliography}
\end{document}